\documentclass{article}
\usepackage{spconf,amsmath,graphicx}
\usepackage{arydshln}
\usepackage{xcolor}
\usepackage{slashbox}
\usepackage{multirow}
\usepackage{amsfonts}
\usepackage{citesort}
\usepackage{booktabs}

\newcommand{\ra}[1]{\renewcommand{\arraystretch}{#1}}

\title{Transcribe-to-Diarize: Neural Speaker Diarization for Unlimited Number of Speakers using End-to-End Speaker-Attributed ASR}
\name{Naoyuki Kanda, Xiong Xiao, Yashesh Gaur, Xiaofei Wang, Zhong Meng, Zhuo Chen, Takuya Yoshioka}
\address{Microsoft, One Microsoft Way, Redmond, WA, USA}
\begin{document}
\ninept
\maketitle
\begin{abstract}
This paper presents Transcribe-to-Diarize, a new approach for neural speaker diarization that uses an end-to-end (E2E) speaker-attributed automatic speech recognition (SA-ASR). The E2E SA-ASR is a joint model that was recently proposed for speaker counting, multi-talker speech recognition, and speaker identification from monaural audio that contains overlapping speech. Although the E2E SA-ASR model originally does not estimate any time-related information, we show that the start and end times of each word can be estimated with sufficient accuracy from the internal state of the E2E SA-ASR by adding a small number of learnable parameters. Similar to the target-speaker voice activity detection (TS-VAD)-based diarization method, the E2E SA-ASR model is applied to estimate speech activity of each speaker while it has the advantages of (i) handling unlimited number of speakers, (ii) leveraging linguistic information for speaker diarization, and (iii) simultaneously generating speaker-attributed transcriptions. Experimental results on the LibriCSS and AMI corpora show that the proposed method achieves significantly better diarization error rate than various existing speaker diarization methods when the number of speakers is unknown, and achieves a comparable performance to TS-VAD when the number of speakers is given in advance. The proposed method simultaneously generates speaker-attributed transcription with state-of-the-art accuracy.
\end{abstract}
\section{Introduction}
\label{sec:intro}

Speaker diarization is a task of recognizing ``{\it who spoke when}'' from audio recordings \cite{park2021review}.
A conventional approach is based on speaker embedding extraction for short segmented audio, 
followed by 
 clustering of the embeddings (sometimes with some constraint regarding the speaker transitions) to attribute the speaker identity to 
each short segment.
Many variants of this approach have been investigated such as 
the methods using agglomerate hierarchical clustering (AHC) \cite{garcia2017speaker}, 
spectral clustering (SC) \cite{park2019auto}, 
and variational Bayesian inference \cite{diez2018speaker,landini2022bayesian}.
While these approaches showed a good performance 
for difficult test conditions \cite{ryant2019second},
they cannot handle overlapped speech \cite{raj2020integration}.
Several extensions  were also proposed to handle overlapping speech, such as
using overlapping detection \cite{bullock2020overlap} and speech separation \cite{xiao2021microsoft}.
However, such extensions typically end up with a combination of multiple heuristic rules,
which is difficult to optimize. 

A neural network-based approach provides more consistent ways to handle the overlapping speech problem by representing the speaker diarization process with a single model.
End-to-end neural speaker diarization (EEND) 
learns a neural network that directly maps an input acoustic feature sequence into a speaker diarization result with permutation-free loss functions \cite{fujita2019end,fujita2019end2}.
Various extensions of EEND were later proposed to cope with an unknown number of speakers \cite{horiguchi2020end,kinoshita2021integrating}.
Region proposal network (RPN)-based speaker diarization \cite{huang2020speaker} uses
a neural network that simultaneously performs speech activity detection, speaker embedding extraction, and
the resegmentation of detected speech regions.
Target-speaker voice activity detection (TS-VAD) \cite{medennikov2020target}
is another approach where the neural network is trained to estimate
speech activities of all the speakers specified by a set of  pre-estimated speaker embeddings.
Of these speaker diarization methods,
TS-VAD 
achieved the state-of-the-art (SOTA) results in 
several diarization tasks \cite{medennikov2020target,raj2020integration} 
including recent international competitions \cite{ryant2020third,wang2021dku}.
On the other hand,
TS-VAD has a limitation that the number of recognizable speakers is bounded by the number of output nodes of the model.

Speaker diarization performance can also be improved by leveraging 
the linguistic information.
For example, the transcription of the input audio provides a strong clue to estimate the utterance boundaries.
Several works were proposed to combine the automatic speech recognition (ASR) with 
speaker diarization, such as 
using the word boundary information from ASR
 \cite{huang2007ibm,silovsky2012incorporation}
or improving the speaker segmentation and clustering based on the information from ASR
 \cite{park2020speaker,xia2021turn}. 
While these works showed promising results,
the ASR and speaker diarization models were separately trained.
Such a combination may not fully utilize the inherent inter-dependency between the speaker diarization and ASR.

With these backgrounds, in this paper,
we present Transcribe-to-Diarize, 
a new speaker diarization approach that uses an
end-to-end (E2E) speaker-attributed automatic speech recognition (SA-ASR) \cite{kanda2020joint} as the backbone.
The E2E SA-ASR 
was originally proposed to recognize ``{\it who spoke what}'' by
jointly performing speaker counting, multi-talker ASR,
and speaker identification 
from monaural audio that possibly contains overlapping speech.
Although the original E2E SA-ASR model does not estimate any information about ``{\it when}'',
in this study, we show that the start and end times of each word can be estimated based on 
the decoder network of the E2E SA-ASR, making the model to recognize ``{\it who spoke when and what}''.
A rule based method for estimating the time information from the attention weights was investigated in our previous work \cite{kanda2020investigation}.
Here we substantially improve the diarization accuracy by introducing a learning based framework. 
In our experiment using the LibriCSS  \cite{chen2020continuous} and AMI \cite{carletta2005ami} corpora, 
we show that the proposed method achieves the SOTA performance in both diarization error rate (DER) and
the concatenated minimum-permutation word error rate (cpWER) \cite{watanabe2020chime} 
for 
the speaker-attributed transcription task.

\section{E2E SA-ASR: review}

\subsection{Overview}
\vspace{-.5em}

The E2E SA-ASR model \cite{kanda2020joint} uses 
 acoustic feature sequence $X\in\mathbb{R}^{f^a\times l^a}$ 
 and a set of 
 speaker profiles $D=\{d_k \in \mathbb{R}^{f^d}|k=1,...,K\}$ as input.
 Here,
 $f^a$ and
$l^a$ are the feature dimension and length of the feature sequence, respectively.
Variable $K$ is the total number of profiles, 
 $d_k$ is the speaker embedding
 (e.g., d-vector \cite{variani2014deep})
of the $k$-th speaker,
and $f^d$ is the dimension of the speaker embedding.
We assume $D$ includes the profiles of all the speakers present in the observed audio.
$K$ can be greater than the actual number of the speakers in the observed audio.

Given $X$ and $D$,
the E2E SA-ASR model
   estimates a
  multi-talker transcription, i.e., word sequence $Y=(y_n\in \{1,...,|\mathcal{V}|\}|n=1,...,N)$
accompanied by
the speaker identity of each token $S=(s_n\in \{1,...,K\}|n=1,...,N)$. 
Here,
$|\mathcal{V}|$  is the size of the vocabulary $\mathcal{V}$, 
$y_n$ is the word index for the $n$-th token,
and $s_n$ is the speaker index for the $n$-th token.
  Following the serialized output training (SOT) framework \cite{kanda2020sot}, 
a multi-talker transcription is represented
as a single sequence $Y$ by concatenating the word sequences of the individual speakers with a special ``speaker change'' symbol $\langle sc\rangle$.
 For example, the reference token sequence to $Y$ for the three-speaker case is given as
$R=\{r^1_{1},..,r^1_{N^1}, \langle sc\rangle, r^2_{1},..,r^2_{N^2}, \langle sc\rangle, r^3_{1},..,r^3_{N^3}, \langle eos\rangle\}$, 
where $r^j_i$ represents the $i$-th token of the $j$-th speaker.
A special symbol $\langle eos\rangle$ is inserted at the end of all transcriptions 
to determine the termination of inference.
Note that this representation 
can be used for overlapping speech of any number of speakers.

\subsection{Model architecture}
\vspace{-.5em}

The E2E SA-ASR model consists of
two attention-based encoder-decoders (AEDs), i.e. an AED for ASR and an AED for speaker identification.
The two AEDs depend on each other, and jointly
estimate $Y$ and $S$ from $X$ and $D$.

The AED for ASR is 
represented as, 
 \begin{align}
 H^{\rm asr} &={\rm AsrEncoder}(X),  \label{eq:enc}  \\
 o_n &= {\rm AsrDecoder}(y_{[1:n-1]}, H^{\rm asr}, \bar{d}_n).  \label{eq:asrout}
 \end{align}
The AsrEncoder module
converts the acoustic feature $X$ 
into a sequence of hidden embeddings $H^{\rm asr} \in \mathbb{R}^{f^h\times l^h}$ for ASR (Eq. \eqref{eq:enc}),
where $f^h$ and $l^h$ are the embedding dimension and 
the length of the embedding sequence, respectively.
The AsrDecoder module then iteratively estimates 
the output distribution $o_n \in \mathbb{R}^{|\mathcal{V}|}$ for $n=1,...,N$ 
 given
previous token estimates $y_{[1:n-1]}$,
$H^{\rm asr}$,
and 
the weighted speaker profile $\bar{d}_n$ (Eq. \eqref{eq:asrout}).
Here, $\bar{d}_n$ is 
calculated in the AED for speaker identification, 
which will be
explained later.
The posterior probability
of token $i$ (i.e. the $i$-th token in $\mathcal{V}$) 
at the $n$-th decoder step 
is represented as
 \begin{align}
Pr(y_n=i|y_{[1:n-1]},s_{[1:n]},X,D) = o_{n,i}, \label{eq:tokenprob}
\end{align}
where $o_{n,i}$ represents
the $i$-th element of $o_n$.

The AED for speaker identification is represented as 
\begin{align}
 H^{spk} &= {\rm SpeakerEncoder}(X),  \label{eq:spkenc} \\
 q_n &= {\rm SpeakerDecoder}(y_{[1:n-1]},H^{\rm spk},H^{\rm asr}), \label{eq:spkquery} \\
\beta_{n,k}&= \frac{\exp(\cos(q_n,d_k))}{\sum_j^K \exp(\cos(q_n,d_j))}, \label{eq:invatt} \\
\bar{d}_n&=\sum_{k=1}^{K}\beta_{n,k}d_k. \label{eq:weighted_prof}
\end{align}
The SpeakerEncoder module converts $X$ into 
a speaker embedding sequence $H^{\rm spk}\in \mathbb{R}^{f^h \times l^h}$
that represents the speaker characteristic of
 $X$ (Eq. \eqref{eq:spkenc}).
The SpeakerDecoder module then 
iteratively estimates 
speaker query $q_n \in \mathbb{R}^{f^d}$ 
for $n=1,...,N$
given $y_{[1:n-1]}$,
$H^{\rm spk}$ and $H^{\rm asr}$ (Eq. \eqref{eq:spkquery}).
A cosine similarity-based attention weight
 $\beta_{n,k}\in \mathbb{R}$
is then calculated 
for all 
 profiles $d_k$ in $D$
given the speaker query $q_n$ (Eq. \eqref{eq:invatt}).
A posterior probability of person $k$ speaking the $n$-th token 
is represented by $\beta_{n,k}$ as
\begin{align}
Pr(s_n=k|y_{[1:n-1]},s_{[1:n-1]},X,D)=\beta_{n,k}. \label{eq:spk-prob}
\end{align}
Finally,
a weighted average of the speaker profiles is calculated as $\bar{d}_n\in \mathbb{R}^{f^d}$ (Eq. \eqref{eq:weighted_prof}) to be fed into the 
AED for ASR
 (Eq. \eqref{eq:asrout}).

The joint posterior probability $Pr(Y,S|X,D)$ can be represented based on Eqs. \eqref{eq:tokenprob} and \eqref{eq:spk-prob} (see \cite{kanda2020joint}).
The model parameters are optimized by maximizing $\log Pr(Y,S|X,D)$ over
training data.

\subsection{E2E SA-ASR based on Transformer}
\vspace{-.5em}

Following \cite{kanda2021end}, 
a transformer-based network architecture is used for the AsrEncoder, AsrDecoder, and SpeakerDecoder modules.
The SpeakerEncoder module 
is based on Res2Net \cite{gao2019res2net}.
Here, we describe only 
the AsrDecoder because it is necessary to explain the proposed method.
Refer to \cite{kanda2021end} for the details of the other modules.

Our AsrDecoder is almost the same as a conventional 
transformer-based decoder \cite{vaswani2017attention} except
for the addition of
 the weighted speaker profile $\bar{d}_n$ 
at the first layer.
The AsrDecoder is represented as 
\begin{align}
&z_{[1:n-1],0}^{\rm asr}=\mathrm{PosEnc}(\mathrm{Embed}(y_{[1:n-1]})),\label{eq:emb}\\
&\bar{z}^{\rm asr}_{n-1,l}=z^{\rm asr}_{n-1,l-1}  \nonumber \\
& \hspace{3mm}+ \mathrm{MHA}_{\rm l}^{\rm asr{\text -}self}(z^{\rm asr}_{n-1,l-1},z_{[1:n-1],l-1}^{\rm asr},z_{[1:n-1],l-1}^{\rm asr}), \label{eq:asr-self}\\
&\bar{\bar{z}}^{\rm asr}_{n-1,l}=\bar{z}^{\rm asr}_{n-1,l} + \mathrm{MHA}_{\rm l}^{\rm asr{\text -}src}(\bar{z}^{\rm asr}_{n-1,l},H^{\rm asr},H^{\rm asr}), \label{eq:asr-src}\\
&z^{\rm asr}_{n-1,l}=
\left\{
\hspace{-2mm}\begin{array}{ll}
\bar{\bar{z}}^{\rm asr}_{n-1,l}+\mathrm{FF}_l^{\rm asr}(\bar{\bar{z}}^{\rm asr}_{n-1,l} + W^{\rm spk}\cdot \bar{d}_n) & \hspace{-3mm}(l=1)  \\
\bar{\bar{z}}^{\rm asr}_{n-1,l}+\mathrm{FF}_l^{\rm asr}(\bar{\bar{z}}^{\rm asr}_{n-1,l})& \hspace{-3mm} (l>1) 
\end{array}
\right. \label{eq:asr-ff}\\
&o_{n} = \mathrm{SoftMax}(W^{o}\cdot z_{n-1,L^{\rm asr}}^{\rm asr} + b^o). \label{eq:asr_out}
\end{align}
Here, $\mathrm{Embed}()$ and $\mathrm{PosEnc}()$ are the embedding
function and absolute positional encoding function \cite{vaswani2017attention}, respectively.
$\mathrm{MHA}^*_l(Q,K,V)$ represents the multi-head attention of the $l$-th layer
\cite{vaswani2017attention} with query $Q$, key $K$, and value $V$ matrices.
$\rm FF_l^{\rm asr}()$ is a position-wise feed forward network in the $l$-th layer.

A token sequence $y_{[1:n-1]}$ is first converted
into a sequence of embedding $z_{[1:n-1],0}^{\rm asr}\in \mathbb{R}^{f^h\times (n-1)}$
(Eq. \eqref{eq:emb}).
For each layer $l$, 
 the self-attention operation 
(Eq. \eqref{eq:asr-self}) 
 and source-target attention operation (Eq. \eqref{eq:asr-src}) are 
 applied.
 Finally, the position-wise feed forward layer is applied to
 calculate the input to the next layer 
$z_{n,l+1}^{\rm asr}$ (Eq. \eqref{eq:asr-ff}).
Here,
$\bar{d}_n$ is added after being multiplied by the weight $W^{spk} \in \mathbb{R}^{f^h \times f^d}$
in the first layer.
Finally, $o_n$
is calculated by applying SoftMax function 
on the final $L^{\rm asr}$-th  layer's output
with
weight $W^o \in \mathbb{R}^{|\mathcal{V}| \times f^h}$ and bias $b^o \in \mathbb{R}^{|\mathcal{V}|}$ 
application
 (Eq. \eqref{eq:asr_out}).

\begin{table*}[t]
\ra{1.0}
  \caption{Comparison of the DER (\%) and cpWER (\%) on the LibriCSS corpus with the monaural setting. Automatic VAD (i.e. not oracle VAD) was used in all systems. The DER including speaker overlapping regions was evaluated with 0 sec of collar. The number with {\bf bold} font is the best result with automatic speaker counting, and
  the number with \underline{underline} is the best result given oracle number of speakers. }
  \label{tab:libricss_summary}
  \vspace{0.5mm}
  \centering
{ \footnotesize
\begin{tabular}{@{}lllllllllllll@{}}
    \toprule
System &&\multirow{2}{*}{\shortstack[l]{Speaker\\counting}}  &&  \multicolumn{7}{c}{DER for different overlap ratio} && cpWER \\ \cmidrule{5-11}
       && &&  0L & 0S & 10 & 20 & 30 & 40 & Avg. &  \\ \midrule
AHC \cite{raj2020integration} && estimated && 16.1 & 12.0 & 16.9 & 23.6 & 28.3 & 33.2 & 22.6 && 36.7$^\star$\\ 
VBx \cite{raj2020integration} && estimated && 14.6 & 11.1 & 14.3 & 21.5  & 25.4 & 31.2 & 20.5 && 33.4$^\star$ \\ 
SC \cite{raj2020integration} && estimated && 10.9  & 9.5 & 13.9 & 18.9 & 23.7 & 27.4 & 18.3 && 31.0$^\star$\\ 
RPN \cite{raj2020integration} && oracle && \underline{4.5} & 9.1 & 8.3 & \underline{6.7} & 11.6 & 14.2 & 9.5 && 27.2$^\star$ \\ 
TS-VAD \cite{raj2020integration} && oracle && 6.0 & \underline{4.6} & \underline{6.6} & 7.3 & 10.3 & 9.5 & \underline{7.6} && 24.4$^\star$ \\ \midrule
SC (ours) && estimated && 9.0  & {\bf 7.9}  & 11.7 & 16.5 & 22.2 & 25.6 & 16.4 && - \\  
SC (ours) && oracle && 7.9   & 7.7  & 11.5 & 16.9 &  20.9 & 25.5 & 16.0 && - \\  \midrule
Transcribe-to-Diarize
&& estimated &&  {\bf 7.2} & 9.5  & {\bf 7.2} & {\bf 7.1} & {\bf 10.2} & {\bf 8.9} & {\bf 8.4} && {\bf 12.9} \\  
Transcribe-to-Diarize
&& oracle && 6.2  & 9.5  & 6.8  & 7.5 & \underline{8.6} & \underline{8.3} & 7.9 && \underline{11.6}\\  \bottomrule
  \end{tabular}
  }
  \\\vspace{0.5mm}{\footnotesize $^\star$ TDNN-F-based hybrid ASR biased by the target-speaker i-vector was applied on top of the diarization results.}
  \vspace{-5mm}
\end{table*}

\begin{figure}[t]
  \centering
  \includegraphics[width=\linewidth]{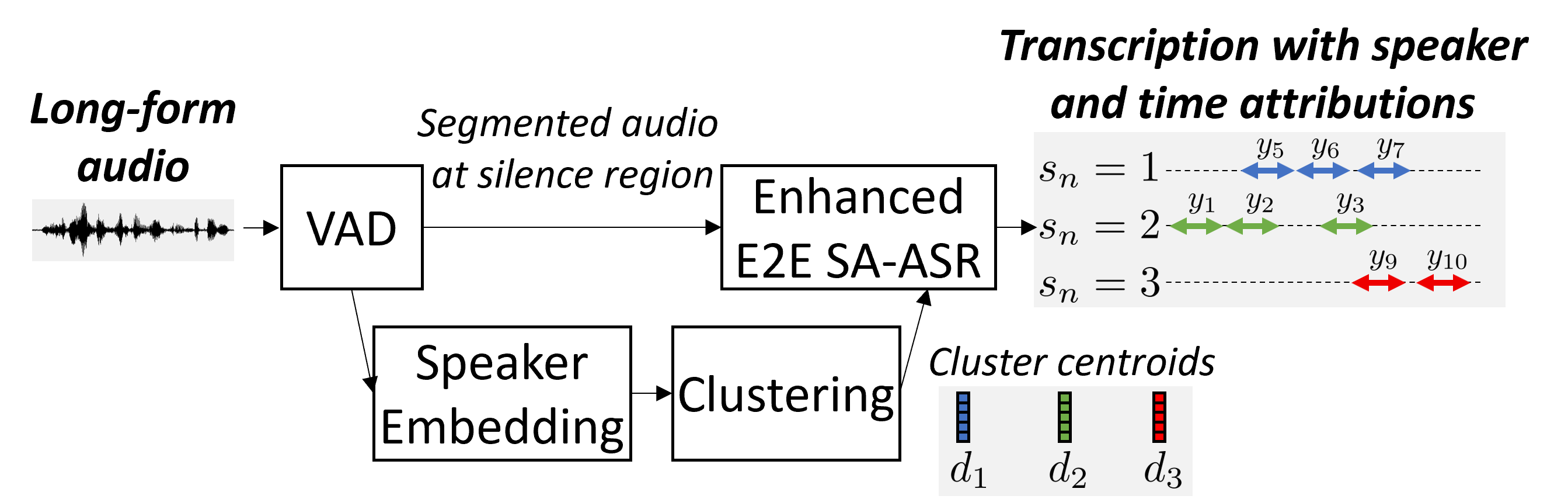}
  \vspace{-7mm}
  \caption{Overview of the proposed approach.}
  \label{fig:overview}
  \vspace{-5mm}
\end{figure}

\section{Speaker diarization using E2E SA-ASR}

\subsection{Procedure overview}
\vspace{-.5em}

The overview of the proposed procedure is shown in Fig. \ref{fig:overview}.
VAD is first applied to the long-form audio to detect silence regions.
Then, speaker embeddings are extracted from uniformly segmented audio with a sliding window.
A conventional clustering algorithm (in our experiment, spectral clustering) is then applied
to obtain the cluster centroids.
Finally, the E2E SA-ASR is applied to each VAD-segmented audio 
  with the cluster centroids as the speaker profiles.
  In this work, the E2E SA-ASR model is extended to generate not only 
  a speaker-attributed transcription but also the start and end times of each token,
  which can be directly translated to the speaker diarization result.
In the evaluation,
detected regions for temporarily close tokens (i.e. tokens apart from each other with less than $M$ sec) with the same speaker identities
are merged to form a single speaker activity region.
 We also exclude abnormal estimations where 
 (i) end\_time - start\_time $\geq$ $N$ sec or (ii) end\_time $<$ start\_time 
 for a single token.
We set $M=2.0$ and $N=2.0$ in our experiment according to the preliminary results.

\subsection{Estimating start and end times from Transformer decoder}
\vspace{-.5em}

In this study, we propose to estimate start and end times of $n$-th estimated token
from the query $\bar{z}_{n-1,l}^{\rm asr}$ and key $H^{\rm asr}$,
which are used in the
 source-target attention (Eq. \eqref{eq:asr-src}),
with a small number of learnable parameters.
 Note that, although there are several prior works
 that conducted the analysis on the source-target attention, 
 we are not aware of
 any prior works that directly estimate the start and end times of each token
 with learnable parameters.
It should also be noted that we can not rely on a conventional force-alignment tool (e.g. \cite{mcauliffe2017montreal})
because the input audio may be including overlapping speech.
 
With the proposed method, the probability distribution of 
start time frame of the $n$-th token 
over the length of $H^{\rm asr}$
is estimated as 
\begin{align}
\alpha_n^{\rm start} = \mathrm{Softmax}(\sum_l \frac{(W_{l}^{\rm s,q} \bar{z}_{n-1,l}^{asr})^\mathsf{T} (W_{l}^{\rm s,k} H^{\rm asr})}{\sqrt{f^{\rm se}}}). \label{eq:start}
\end{align}
Here, $f^{\rm se}$ is the dimension of the subspace to estimate the start time frame of each token.
The terms $W_{l}^{\rm s,q}\in\mathbb{R}^{f^{\rm se}\times f^h}$ and 
$W_{l}^{\rm s,k}\in\mathbb{R}^{f^{\rm se}\times f^h}$ are the affine transforms to map the 
query and key to the subspace, respectively.
The resultant $\alpha_n^{\rm start} \in \mathbb{R}^{l^h}$ is the scaled dot-product attention accumulated for all layers, and it can be regarded as the probability 
distribution of the start time frame of $n$-th token 
 over the length of embeddings $H^{\rm asr}$.
Similarly, the probability distribution of the end time frame of the $n$-th token, represented by $\alpha_n^{\rm end} \in \mathbb{R}^{l^h}$,  is estimated 
by replacing 
 $W_{l}^{\rm s,q}$ and $W_{l}^{\rm s,k}$
 of Eq \eqref{eq:start}
 with 
 $W_{l}^{\rm e,q}\in\mathbb{R}^{f^{\rm se}\times f^h}$ and 
$W_{l}^{\rm e,k}\in\mathbb{R}^{f^{\rm se}\times f^h}$, respectively.

The parameters 
$W_{l}^{\rm s,q}$,
$W_{l}^{\rm s,k}$,
$W_{l}^{\rm e,q}$, and
$W_{l}^{\rm e,k}$ are learned from training data that 
includes the reference start and end time indices 
on the embedding length of $H^{\rm asr}$.
In this paper, we apply a cross entropy (CE) objective function
on the estimation of $\alpha_n^{\rm start}$
and $\alpha_n^{\rm end}$ on every token except special tokens  $\langle sc\rangle$ and  $\langle eos\rangle$.
We
perform the multi-task training with 
 the objective function of the original E2E SA-ASR model 
and the objective function of the start-end time estimation
with an equal weight to each objective function.
In the inference, frames with the maximum value on $\alpha_n^{\rm start}$ and $\alpha_n^{\rm end}$
are selected as the start and end frames for the $n$-th token, respectively.

\begin{table*}[t]
\ra{1.1}
  \caption{Comparison of the DER (\%) and cpWER (\%) on the AMI corpus. The number of speakers was estimated in all systems. The DER including speaker overlapping regions was evaluated with 0 sec of collar based on the reference boundary determined in \cite{landini2022bayesian}. 
  SER: speaker error, Miss: miss error, FA: false alarm. DER = SER + Miss + FA. }
  \label{tab:ami_summary}
  \vspace{0.5mm}
  \centering
{\footnotesize
\begin{tabular}{@{}llllllll@{}}
    \toprule
Audio & System & VAD  &  \multicolumn{2}{c}{dev} && \multicolumn{2}{c}{eval} \\ \cmidrule{4-5} \cmidrule{7-8} 
&       &               &  \multicolumn{1}{c}{SER / Miss / FA / DER} & cpWER && \multicolumn{1}{c}{SER / Miss / FA / DER} & cpWER  \\ \midrule
IHM-MIX & AHC \cite{landini2022bayesian} & oracle$^\dagger$ & 6.16 / 13.45 / 0.00 / 19.61  & - && 6.87 / 14.56 / 0.00 / 21.43 & - \\
IHM-MIX & VBx \cite{landini2022bayesian} & oracle$^\dagger$ & 2.88 / 13.45 / 0.00 / 16.33  & - && 4.43 / 14.56 / 0.00 / 18.99 & - \\ \midrule
IHM-MIX & SC & automatic & 3.37 / 14.89 / 9.67 / 27.93 & 23.1$^\ddagger$ &&  3.45 / 16.34 / 9.53 / 29.32  & 23.4$^\ddagger$\\
IHM-MIX & 
Transcribe-to-Diarize
& automatic        &  3.05 / 11.46 / 9.00 / 23.51 & 15.9 && 2.47 / 14.24 / 7.72 / 24.43 & 16.4 \\ 
IHM-MIX & 
Transcribe-to-Diarize
& oracle$^\dagger$ &  2.83 / $\:$ 9.69 / 3.46 / 15.98 & 16.3 && 1.78 / 11.71 / 3.10 / 16.58 & 15.1 \\ \midrule
SDM & SC & automatic  &  3.50 / 21.93 / 4.54 / 29.97 & 28.6$^\ddagger$ && 3.69 / 24.84 / 4.14 / 32.68 & 30.3$^\ddagger$\\
    SDM & 
    Transcribe-to-Diarize
    & automatic & 3.48 / 15.93 / 7.17 / 26.58 & 22.6 && 2.86 / 19.20 / 6.07 / 28.12 & 24.9 \\ 
SDM & 
Transcribe-to-Diarize
& oracle$^\dagger$  & 3.38 / 10.62 / 3.28 / 17.27  & 21.5 && 2.69 / 12.82 / 3.04 / 18.54  & 22.2 \\ \bottomrule
  \end{tabular}
  }\\
{\footnotesize \hspace{-53mm} $^\dagger$ Reference boundary information was used to segment the audio at each silence region. }\\
{\footnotesize \hspace{5.9mm} $^\ddagger$ Single-talker Conformer-based ASR pre-trained by 75K-data and fine-tuned by AMI \cite{kanda2021comparative} 
  was used on top of the speaker diarization result.}\\
  \vspace{-5mm}
\end{table*}

\section{Evaluation Results}

We evaluated the proposed method with the LibriCSS corpus \cite{chen2020continuous} and
the AMI meeting corpus \cite{carletta2005ami}.
We used DER as the primary performance metric. 
We also used
the cpWER \cite{watanabe2020chime} 
for the evaluation
of speaker-attributed transcription.

\subsection{Evaluation on the LibriCSS corpus}
\vspace{-.3em}
\subsubsection{Experimental settings}
\vspace{-.5em}

The LibriCSS corpus \cite{chen2020continuous} is a set of 8-speaker recordings made by 
playing back ``test\_clean" of LibriSpeech in a real meeting room.
The recordings are 10 hours long in total, and 
they are categorized by the speaker overlap ratio from 0\% to 40\%.
We used the first channel of the 7-ch microphone array recordings in this experiment.

We used the model architecture described in \cite{kanda2021comparative}.
The AsrEncoder consisted of 
2 convolution layers that subsamples the time frames by a factor of 4,
followed by
18 Conformer \cite{gulati2020conformer} layers.
The AsrDecoder consisted of
6 layers
and 16k subwords 
were used as a recognition unit.
The SpeakerEncoder 
was based on Res2Net \cite{gao2019res2net}
and designed to be the same with that of the speaker profile extractor.
Finally, SpeakerDecoder consisted of 2 transformer layers.
 We used a 80-dim log mel filterbank extracted every 10 msec as the input feature,
and the Res2Net-based d-vector extractor \cite{xiao2021microsoft} trained by 
VoxCeleb corpora \cite{nagrani2017voxceleb,chung2018voxceleb2}
 was used to extract a 128-dim speaker embedding. 
 We set $f^{\rm se}=64$ for the start and end-time estimation.
 See \cite{kanda2021comparative} for more details of the model architecture.

We 
used a similar multi-speaker training data set to the one used in \cite{kanda2020investigation}
 except that we newly introduced 
a small amount of training samples with no overlap between the speaker activities.
The training data were generated by mixing 1 to 5 utterances of 1 to 5 speakers from LibriSpeech
with random delays being applied to each utterance, where 90\% of the delay was designed to have
speaker overlaps 
while 10\% of the delay was designed to have no speaker overlaps with 0 to 1 sec of intermediate silence.
 Randomly generated room impulse responses and noise were also added to simulate the reverberant recordings.
We used the word alignment information on the original LibriSpeech utterances (i.e. the ones before mixing) generated with the Montreal Forced Aligner \cite{mcauliffe2017montreal}. 
If one word consists of multiple subwords, we divided the duration of such a word
by the number of subwords to determine the start and end times of each subword.
We initialized the ASR block by the model trained with SpecAugment as described in \cite{kanda2021end},
 and 
 performed
 160k iterations of training 
 based only with $\log Pr(Y,S|X,D)$ 
 with a mini-batch of 6,000 frames with 8 GPUs.
 Noam learning rate schedule with peak learning rate of 0.0001 after 10k iterations was used.
We then reset the learning rate schedule, and performed 
further 80k iterations of training 
 with the training objective function that includes the CE objective function 
 for the start and end times of each token.

In the evaluation,
we first 
applied the WebRTC VAD\footnote{
https://github.com/wiseman/py-webrtcvad}
with the least aggressive setting, 
and extracted the d-vector from speech region based on the
1.5 sec of sliding window with 0.75 sec of window shift.
Then, we 
applied the speaker counting and clustering based on the normalized maximum eigengap-based
spectral clustering (NME-SC) \cite{park2019auto}.
Then,
we cut the audio into a short segment at the middle of silence region detected by 
WebRTC VAD. We split the audio when the duration of the audio was longer than the 20 sec. 
We then ran 
the E2E SA-ASR for each segmented audio 
with the average speaker embeddings from the speaker cluster generated by the NME-SC.

\subsubsection{Evaluation Results}
\vspace{-.5em}

Table \ref{tab:libricss_summary} shows the result on the LibriCSS corpus with various diarization 
methods. With the estimated number of speakers,
our proposed method achieved a significantly better average DER (8.4\%)
than any other speaker diarization techniques (16.0\% to 22.6\%).
We also confirmed that the proposed method achieved
the DER of 7.9\% when the number of speakers was known, which was close to 
the strong result by TS-VAD (7.6\%) that was specifically trained for the 8-speaker condition.
We observed that the proposed method was especially good for the inputs with high overlap ratio,
such as 30\% to 40\%.
It should be also noted that the cpWER by the proposed method (11.6\% with 
the oracle number of speakers and 12.9\% with the estimated number of speakers)
were significantly better than the prior works,
and they are the SOTA results
for the monaural setting of LibriCSS with no prior knowledge on speakers.



\subsection{Evaluation on the AMI corpus}
\vspace{-.3em}
\subsubsection{Experimental settings}
\vspace{-.5em}

We also evaluated the proposed method with
 the AMI meeting corpus \cite{carletta2005ami},
 which is a set of real meeting recordings of four participants.
For the evaluation, 
we used  single distant microphone (SDM) recordings
or the mixture of independent headset microphones, called IHM-MIX.
 We used scripts of the Kaldi toolkit \cite{povey2011kaldi} to
partition
the recordings into training, development, and evaluation sets.
The total durations of the three sets were
  80.2 hours, 9.7 hours, and 9.1 hours, respectively.
  
We 
initialized the model with
a well-trained E2E SA-ASR model 
based on the 75 thousand hours of ASR training data,  
 VoxCeleb corpus, and AMI-training data, 
the detail of which is described in \cite{kanda2021comparative}. 
We performed 2,500 training iterations by using the AMI training corpus
 with a mini-batch of 6,000 frames with 8 GPUs
and a linear decay learning rate schedule with peak learning rate of 0.0001.
 We used the word boundary information obtained from the reference annotations of the AMI corpus. 
Unlike the experiment on the LibriCSS, 
 we 
 updated only 
 $W_{l}^{\rm s,q}$,
$W_{l}^{\rm s,k}$,
$W_{l}^{\rm e,q}$, and
$W_{l}^{\rm e,k}$ by freezing other pre-trained parameters because overfitting was observed otherwise 
in our preliminary experiment.

\subsubsection{Evaluation Results}
\vspace{-.5em}

The evaluation results are shown in Table \ref{tab:ami_summary}. 
The proposed method achieved significantly better DERs for
both SDM and IHM-MIX conditions.
Especially, we observed significant improvements in the miss error rate, which 
indicates the effectiveness of the proposed method in the speaker overlapping regions.
The proposed model, pre-trained on a large-scale data \cite{kanda2021large,kanda2021comparative},
simultaneously
achieved the SOTA cpWER on the AMI dataset among fully automated monaural SA-ASR systems.

\section{Conclusion}
This paper presented the new approach for speaker diarization
using
the E2E SA-ASR model. 
In  the  experiment with the LibriCSS corpus and the AMI meeting corpus,
the  proposed  method  achieved  significantly  better DER 
over various speaker diarization methods under the condition of the speaker number being unknown 
while
achieving almost the same DER as TS-VAD when the oracle speaker number is available.

\bibliographystyle{IEEEtran}
\bibliography{mybib}

\end{document}